\documentclass[11pt]{article}
\usepackage{float}
\usepackage{mymacros}
\usepackage{chngcntr}
\usepackage{verbatim}
\usepackage{amsthm}
\usepackage{graphics}
\usepackage{graphicx}
\usepackage{xcolor}
\usepackage{amsmath}
\usepackage{amssymb}
\usepackage{mathrsfs}
\usepackage{bbold}
\usepackage{cases}
\usepackage{epsfig}
\usepackage{epstopdf}
\usepackage{hyperref}
\usepackage{multirow}
\usepackage{amsfonts}
\usepackage{hyperref}

\hypersetup{
colorlinks=true,
linkcolor=blue,
filecolor=blue,
urlcolor=blue,
citecolor=blue,
}

\title{\boldmath Bound on Lyapunov exponent in Einstein-Maxwell-Dilaton-Axion black holes}

\author{Chengye Yu, $^{a}$\footnote{E-mail: chengyeyu1@hotmail.com}}
\author{Deyou Chen, $^{a}$\footnote{E-mail: deyouchen@hotmail.com}}
\author{Chuanhong Gao$^{a}$\footnote{E-mail: chuanhonggao@hotmail.com}}

\makeatletter

\@addtoreset{equation}{section}
\makeatother

\affiliation{$^{a}$School of Science, Xihua University, Chengdu 610039, China}

\abstract{In this paper, we investigate the influence of the angular momentum of a charged particle around non-extremal and extremal Einstein-Maxwell-Dilaton-Axion black holes on the Lyapunov exponent. The angular momentum's ranges and spatial regions where the bound of the exponent is violated are found for certain values of the rotation parameter and dilatonic constant of the black holes. This violation always exists when the rotation parameter is large enough and the rotation directions of the particle is opposite to those of the black holes. The spatial regions outside the extermal black hole for the violation is relatively large. In the near-horizon regions of the extremal black holes, the violation depends on the rotation directions of the black holes and particle, and does not depend on the value of the angular momentum.}


\begin{document}
\maketitle
\section{Introduction }

Chaos is an important physical phenomenon in nonlinear dynamic systems. In classical chaos, trajectories of dynamics sensitively depend on initial conditions. A small change of the conditions leads to that the trajectories deviate exponentially from their initial locations. Its sensitivity is characterized by a Lyapunov exponent. However, it is not easy to study the chaos in quantum systems. Recently, people have proposed that the chaos can be efficiently diagnosed from a holographic approach by using out-of-time-order correlators (OTOCs) \cite{2015JHEP...05..132S,PhysRevLett.115.131603, 2014JHEP...03..067S,2014JHEP...12..046S, 2015JHEP...03..051R}. In holographic systems at finite temperature, the OTOC is 

\begin{eqnarray}
\frac{<V(0)W(t,\vec{x})V(0)W(t,\vec{x})>}{<V(0)V(0)>W(t,\vec{x})W(t,\vec{x})>} = 1- \varepsilon_{\Delta _V \Delta_W}exp\left[\lambda _L (t-t_{\star}-\frac{|\vec{x}|}{v_B})\right],
\label{eq1.2}
\end{eqnarray}

\noindent where $\varepsilon_{\Delta _V \Delta_W}$ is a multiplicative factor and contains information about the operators $V$ and $W$, $\lambda _L$ is a Lyapunov exponent, $t_{\star}$ is a scrambling time, and $v_B$ is a butterfly velocity. 

In the seminal work \cite{2016JHEP...08..106M}, Maldacena, Shenker and Stanford conjectured that there is a universal bound on the exponent of chaos in thermal quantum systems with a large number of degrees of freedom, 

\begin{eqnarray}
\lambda \leq \frac{2 \pi k_B T}{\hbar},
\label{eq1.1}
\end{eqnarray}

\noindent where $T$ is the temperature of the system. This conjecture relies on two assumptions. The first one is that certain time-ordered correlation functions approximately factorize. Another assumption is that there is a large hierarchy between the dissipation time and scrambling time. This work provides a cornerstone for the development of the AdS/CFT correspondence, and promotes the research on black holes. After the conjecture was put forward, it attracted great attention and was confirmed by a lot of work \cite{2018JCAP...12..036L,2020arXiv200707744G, 2015arXiv151008870B,2016NuPhB.911..191B,2017NuPhB.921..727B, 2018PhRvE..98a2216T, 2018PhRvL.120t1604D, 2019PhLB..788..486D, 2018PhRvD..98h6007H, 2020arXiv200705949T,2019JHEP...04..082G,2019PhRvD..99d6011B,2018arXiv181101079C,2019JHEP...04..056A,2019JHEP...01..048G,2019JHEP...07..099T,2020PhRvB.101q4313C,2020PhRvD.102l4047D,2021PhRvD.104j1901B,2016JHEP...04..001P,2016PhRvD..94j6002M, 2016arXiv160601857M, 2019PhRvD.100d1901H,2019JHEP...03..079B,PhysRevD.100.125005,2017EPJC...77..208W,2016JHEP...09..082C,2018arXiv180609993L,2011IJTP...50..106S,2014PhRvD..89h6011M,2017PhLB..768..288L,2018PhRvD..97f6029W,2018PhRvD..97j6018W,2016JHEP...11..032A}. It was found that the bound is saturated in the Sachdev-Ye-Kitaev (SYK) model, which provides an important basis for the theory dual to gravity \cite{2018JHEP...05..183K}. In the study of particle motions near black holes, when electromagnetic or scalar forces on a particle are large enough, the particle can be very close to the event horizons without falling into them. Based on this, Hashimoto and Tanahashi found that the value of the exponent is independent on the external forces and the particle mass, and obeys an inequality $\lambda \leq \kappa $, where $\kappa$ is the surface gravity of the black hole \cite{2017PhRvD..95b4007H}. From the relation between the temperature and surface gravity of the black hole, this inequality is identical to Eq. (\ref{eq1.1}).

However, as the authors said in \cite{2016JHEP...08..106M}, there are cases where the bound on the exponent (chaos bound) is not applicable \cite{2016arXiv160106164F,2016JHEP...10..069P,2018PhRvD..98l4001Z,2017arXiv171104768S,2019JHEP...12..150C,2018arXiv181210073P,2019JHEP...10..077D,2021PhRvD.104d6020L,2022PhRvD.105b6006K,2021arXiv211106089L,2022arXiv220307298G}. For example, the string corrections to the bound and the cases that does not meet the assumptions. In the classical limit of the SYK model, the linear relation of the exponent dependent on temperature was found \cite{2017arXiv171104768S}. Its slope is different in parameters from that obtained in the quantum case. In the research of the chaos in anti-de Sitter (AdS) spacetimes, the exponent for the motion of classical closed strings was modified as $\lambda =2\pi Tn$ for winding strings in the bulk, where $n$ is the winding number of the string \cite{2019JHEP...12..150C}. Sub-leading terms in near-horizon expansions have an important influence on the exponent's value. When a charged particle is in equilibrium outside an event horizon of a black hole by a Lorentz force, one can adjust the charge mass ratio of the particle to let the particle close to the event horizon. In the near-horizon regions, the bound is violated by the Einstein-Maxwell-Dilaton, Einstein-Born-Infeld and Einstein-Gauss-Bonnet Maxwell black holes, and satisfied by Reissner-Nordstr$\ddot{o}$m (RN) and RN-AdS black holes \cite{2018PhRvD..98l4001Z}. In \cite{2018PhRvD..98l4001Z}, the influence of the angular momentum of the particle was neglected. In fact, the angular momentum plays an important role in the exponent. When this influence is considered, Kan and Gwak studied the bound via the effective potential of the particle \cite{2022PhRvD.105b6006K}. The violation for the bound was found for the specific values of the black hole's parameter. The exponent can also be obtained by the matrix method. Using this method, Lei and Ge found that the bound in the near-horizon regions of the RN and RN-AdS black holes is violated when the angular momentum of the particle and the charge of the black holes are large enough \cite{2021arXiv211106089L}. 

In this paper, we investigate the influence of the angular momentum of a charged particle around extremal and non-extremal Einstein-Maxwell-Dilaton-Axion (EMDA) black holes on the Lyapunov exponent, and find the angular momentum's ranges and spatial regions where the bound is violated. The exponent is derived by the effective potential of the particle and affected by the angular momentum. We first investigate the exponent at a certain distance from the event horizons by numerical calculations. Then the exponent in the near-horizon regions is discussed. In the investigation, the same and opposite rotation directions of the particle and black holes have different influences, which are considered.

The rest is organized as follows. In the next section, we review the EMDA black holes and derive the Lyapunov exponent by the effective potential. An auxiliary field is introduced and a static gauge is used. In Section \ref{b}, we investigate the influence of the angular momentum of the particle around the non-extremal and extremal EMDA black holes on the exponent, and find the spatial regions where the bound is violated. The last section is devoted to our conclusion and discussion.

\section{EMDA black holes and Lyapunov exponent}\label{a}

The EMDA black hole is a solution of field equations arising in the low energy heterotic string field theory and describes a rotating charged spacetime. From the action,

\begin{eqnarray}
S=\int d^{4}x\sqrt{-g}\left[R-2g^{\mu \nu }\partial _{\mu }\phi \partial _{\nu }\phi -\frac{1}{2}e^{4\phi }g^{\mu \nu }\partial _{\mu }\kappa_0 \partial _{\nu }\kappa_0 -e^{-2\phi }F_{\mu \nu }F^{\mu \nu }-\kappa_0 F_{\mu \nu }\check{F}^{\mu \nu }\right],
\label{eq2.1}
\end{eqnarray}

\noindent where $R$ is a scalar Riemann curvature, $F_{\mu \nu }$ is an electromagnetic tensor field and its dual is $\check{F}_{\mu \nu }=-\frac{1}{2}\sqrt{-g}\epsilon _{\mu \nu \alpha \beta }F^{\alpha \beta }$. $\phi$ is a dilaton field, and $\kappa_{0}$ is an Axion scalar field dual to the three-index anti-symmetric tensor field $H=-exp(4\phi)*d\kappa_{0}/4$. From the action, the solutions of rotating black holes were obtained \cite{1992PhRvL..69.1006S,PhysRevLett.74.1276,2008PhRvD..78d4007H}. In \cite{PhysRevLett.74.1276}, Garcia, Galtsov and Kechkin got the EMDA black hole solution, which is given by

\begin{eqnarray}
ds^{2}=-\frac{\Delta }{\Sigma }\left(dt-a\sin^{2}\theta d\phi \right)^{2}+\frac{\sin^{2}\theta }{\Sigma }\left[adt-(\Sigma +a^{2}\sin^{2}\theta )d\phi \right]^{2}+\frac{\Sigma }{\Delta }dr^{2}+\Sigma d\theta ^{2},
\label{eq2.2}	
\end{eqnarray}

\noindent with an electromagnetic potential

\begin{eqnarray}
A_{\mu}dX^{\mu}=\frac{Qr}{\Sigma }dt-\frac{aQr\sin^{2}\theta }{\Sigma }d\phi,
\label{eq2.3}
\end{eqnarray}

\noindent where

\begin{eqnarray}
\Delta &=&r^{2}-2M_{0}r+a^{2}=(r-r_{+})(r-r_{-}),\nonumber\\
\Sigma &=&r^{2}+2br+a^{2}\cos^{2}\theta ,\nonumber\\
M_{0}&=&M - b=M- \frac{Q^{2}}{2M}.
\label{eq2.4}
\end{eqnarray}

\noindent $r_+$ and $r_-$ are the event and inner horizons, respectively. $a$ is a rotation parameter of the black hole, $M$ is the ADM mass and $Q$ is the charge. $b$ is a dilatonic constant and relates to the ADM mass and charge by $b= \frac{Q^{2}}{2M}$. When $a = 0$, the metric describes a charged, non-rotating dilatonic black hole. When $b = 0$, the metric is reduced to the Kerr metric. The event (inner) horizons $r_+ (r_-)$ and surface gravity are  

\begin{eqnarray}
r_{\pm} = M_{0} \pm \sqrt{M_0^2 -a^2}, \quad\quad
\kappa =\frac{r_{+}-r_{-}}{2r_{+}(r_{+}+r_{-}+2b)},
\label{eq2.5}
\end{eqnarray}

\noindent respectively. When the inner and event horizons coincide with each other, the black hole is extremal and the surface gravity is zero.

We consider a charged particle with mass $m$ and charge $q$ moving around the EMDA black hole. Then the action of the particle is \cite{2022PhRvD.105b6006K}

\begin{eqnarray}
S=\int ds\left[\frac{1}{2e(X (s))}g_{\mu \nu }(X(s))\dot{X}^{\mu }(s)\dot{X}^{\nu }(s)-\frac{e(X(s))}{2}m^{2}-qA_{\mu}(X(s))\dot{X^{\mu }}(s)\right].
\label{eq2.6}
\end{eqnarray}

\noindent In the above equation, $e$ is an auxiliary field and $s$ is adopted to parameter the geodesic of the particle. Without loss of generality, we use a static gauge and let $s$ be equal to the time $t$. We focus our attention on the motion of the particle in the equatorial plane of the black hole, where $\theta = \frac{\pi}{2}$. From the metric ({\ref{eq2.2}}) and action ({\ref{eq2.6}}), the Lagrangian is 

\begin{eqnarray}
\mathcal{L}&=&\frac{1}{2e}\left[\frac{(r^{2}+2br)\dot{r}^{2}}{\Delta }+\frac{2a[\Delta -(r^{2}+2br+a^{2})]\dot{\phi }}{r^{2}+2br}+\frac{[(r^{2}+2br+a^{2})^{2}-\Delta a^{2}]\dot{\phi }^{2}}{r^{2}+2br}\right.\nonumber\\
&-&\left.\frac{\Delta -a^{2}}{r^{2}+2br} \right] - \frac{e}{2}m^{2}-\frac{qQ}{r+2b}+\frac{aqQ}{r+2b}\dot{\phi },
\label{eq2.7}
\end{eqnarray}

\noindent where $\dot{\phi }$ appears, and the corresponding angular momentum is

\begin{eqnarray}
L=\frac{\partial \mathcal{L}}{\partial \dot{\phi }}=\frac{a[\Delta -(r^{2}+2br+a^{2})]}{e(r^{2}+2br)}+\frac{[(r^{2}+2br+a^{2})^{2}-\Delta a^{2}]\dot{\phi }}{e(r^{2}+2br)}+\frac{aqQ}{r+2b}.
\label{eq2.8}
\end{eqnarray}

\noindent The equation of motion of the auxiliary field satisfies $-e^{2}m^{2}=\dot{X}^{\mu }\dot{X}_{\mu }$.  The auxiliary field is solved and takes the form $e =(r+2b)\sqrt{r}\sqrt{\frac{\Delta ^{2}-[(r^{2}+2br+a^{2})^{2}-\Delta a^{2}]\dot{r}^{2}}{\alpha }}$. Then the effective Lagrangian of the particle is 

\begin{eqnarray}
\mathcal{L}_{eff}=\mathcal{L}-L\dot{\phi } =-\sqrt{h-\frac{\dot{r}^{2}}{f}}-\Phi,
\label{eq2.11}
\end{eqnarray}

\noindent where

\begin{eqnarray}
h &=& \frac{r\alpha }{\chi ^{2}},\quad \quad 
f = \frac{\chi \Delta ^{2}}{r\alpha }, \quad  \quad 
\chi =(r^{2}+2br+a^{2})^{2}-\Delta a^{2},\nonumber\\
\Phi &=&\frac{a[aqQ-L(r+2b)][\Delta -(r^{2}+2br+a^{2})]}{(r+2b)\chi }+\frac{qQ}{r+2b} ,\nonumber\\
\alpha &=& \Delta [m^{2}(r+2b)[(r^{2}+2br+a^{2})^{2}-\Delta a^{2}]+r[aqQ-L(r+2b)]^{2}].
\label{eq2.12}
\end{eqnarray}

\noindent When the particle moves slowly around a local maximum of a potential and its velocity obeys $\dot{r}\ll 1$. Eq. (\ref{eq2.11}) is expanded and rewritten as follows

\begin{eqnarray}
\mathcal{L}_{eff}=\frac{\dot{r}^{2}}{2\sqrt{h}f}-V_{eff}(r)+\mathcal{O}(\dot{r}^{4}),
\label{eq2.13}
\end{eqnarray}

\noindent where

\begin{eqnarray}
\quad\quad V_{eff}(r)=\sqrt{h}+\Phi ,
\label{eq2.14}
\end{eqnarray}

\noindent is an effective potential, $\mathcal{O}(\dot{r}^{4})$ contains higher order terms of $\dot{r}$ and is neglected. The local maximum of the potential is obtained at a location $r_0$ and determined by $V_{eff}^{'}(r) = 0$, where "$\prime$" represents a derivative in term of $r$ and

\begin{eqnarray}
V_{eff}^{'}(r)=\frac{{h}'}{2\sqrt{h}}+{\Phi }'.
\label{eq2.15}
\end{eqnarray}

\noindent At this location, we introduce a small perturbation $\epsilon (s)$ to let $r(s )=r_{0}+\epsilon (s)$. Then the effective Lagrangian is  

\begin{eqnarray}
\mathcal{L}_{eff}=\frac{1}{2\sqrt{h}f}(\dot{\epsilon }^{2}+\lambda ^{2}\epsilon^{2}).
\label{eq2.16}
\end{eqnarray}

\noindent In the above derivation, the constant and higher-order terms were neglected. $\lambda$ is defined as a Lyapunov exponent and given by

\begin{eqnarray}
\lambda ^{2}=\left. -\sqrt{h}f{V}''_{eff}(r)\right|_{r=r_{0}},
\label{eq2.17}
\end{eqnarray}

\noindent The stability of the system is determined by the exponent. When $\lambda^{2}>0$, the system is unstable and a chaos appears. $\lambda^{2} < 0$ corresponds to the stable system, and $\lambda^{2} = 0$ reflects that the system is marginal. It is not difficult to get $\left.\chi \right|_{r=r_{0}}> 0$ and $\left.\sqrt{h}f\right|_{r=r_{0}}> 0$. Then the sign is determined by the value of ${V}''_{eff}(r)$. Due to the appearance of the angular momentum of the particle, we need to consider the influence of the angular momentum on the exponent when the bound of the exponent is discussed.

\section{Bound on Lyapunov exponent and its violation in EMDA black holes}\label{b}

In this section, we use Eq. (\ref{eq2.17}) to investigate the influence of the angular momentum of the particle around the non-extremal and extremal EMDA black holes on the exponent, and find the angular momentum's ranges and spatial regions where the bound is violated.

\subsection{Lyapunov exponent in non-extremal EMDA black holes}

We first investigate the influence of the angular momentum of the particle around a non-extremal EMDA black hole on the exponent. In \cite{2018PhRvD..98l4001Z}, the authors found that when the charge mass ratio of the particle is large, the particle is in equilibrium near the horizons. In this subsection, we order $M=1$, $m=1$, $q=15$. When $b=\frac{1}{3}$, we use Eq. (\ref{eq2.5}) and get the location of the horizon and the value of the surface gravity. There are $r_{+}=1.26091$ and $\kappa ^2=0.0563241$ when $a=\frac{2}{7}$. When $a=\frac{1}{3}$, we obtain $r_{+}=1.24402$ and $\kappa ^2=0.0538476$. $a=\frac{2}{5}$ yields $r_{+}=1.20000$ and $\kappa ^2=0.0493827$. $a=\frac{1}{2}$ yields $r_{+}=1.10763$ and $\kappa ^2=0.0396232$. It is found from Eq. (\ref{eq2.15}) that different values of the angular momentum and rotation parameter lead to different locations corresponding to the local maximum of the effective potential. Their relations are listed in Table \ref{t.1}. The positive sign in front of the angular momentum indicates that the particle and black hole rotate in the same directions, and the negative sign indicates that they rotate in the opposite directions.

\begin{table}[h]
	\centering
	\begin{tabular}{ccccccccc}
		\hline
		&L&-30&-20&-10&0&10&20&30\\
		\hline
		\multirow{4}{*}{$r_{0}$}&$a=\frac{2}{7}$&2.06440&1.92212&1.67266&1.29859&1.38420&1.49575&1.56067\\
		\cline{2-9}
		&$a=\frac{1}{3}$&2.09597&1.94687&1.68472&1.28002&1.34305&1.44363&1.50221\\
		\cline{2-9}
		&$a=\frac{2}{5}$&2.13829&1.97951&1.69912&1.24754&1.27777&1.36285&1.41242\\
		\cline{2-9}
		&$a=\frac{1}{2}$&2.19796&2.02433&1.71543&1.18062&1.15620&1.21716&1.25265\\
		\hline
	\end{tabular}
	\caption{Locations of equilibrium orbits of the particle around the non-extreme EMDA black hole are gotten when $b=\frac{1}{3}$.}
	\label{t.1}
\end{table}

\begin{figure}[H]
	\centering
	\includegraphics[scale=1.5]{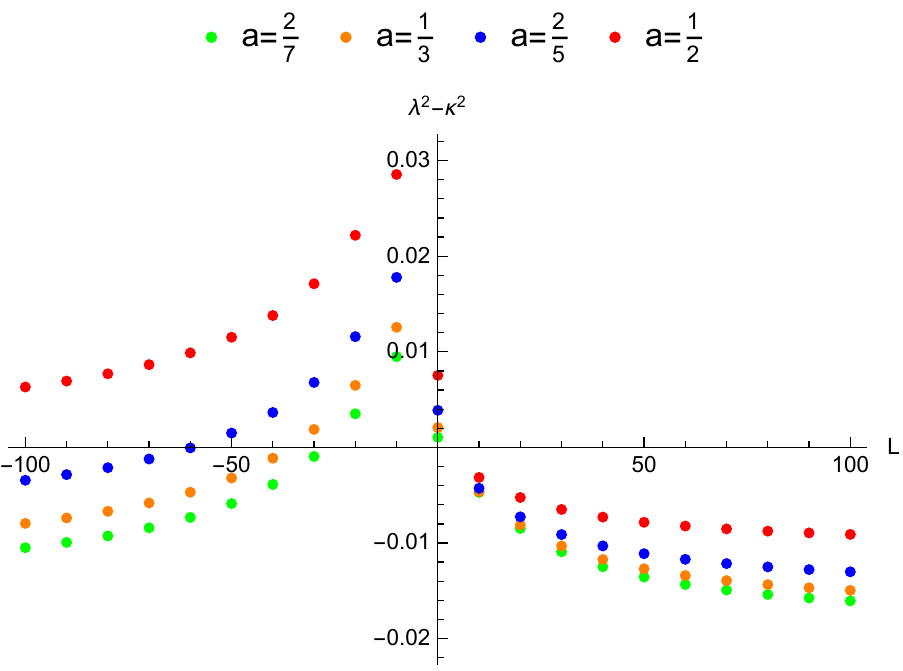}
	\caption{The influence of the angular momentum of the particle around the non-extremal EMDA black hole on the Lyapunov exponent, where $b=\frac{1}{3}$. The bound on the Lyapunov exponent is violated in the range $-27.48<L<0.67$ (the corresponding spatial region is $1.01494866r_{+}<r_{0}<1.60347830r_{+}$) when $a=\frac{2}{7}$, violated in the range $-35.76<L<1.16 (1.01322326r_{+}<r_{0}<1.73241588r_{+})$ when $a=\frac{1}{3}$, violated in the range $-59.63<L<1.84 (1.01115833r_{+}<r_{0}<1.96475153r_{+})$ when $a=\frac{2}{5}$, and violated in the range $L<2.90 (1.00838728r_{+}<r_{0}<2.52098625r_{+})$ when $a=\frac{1}{2}$.}
	\label{fig:1}
\end{figure}

Using Eqs. (\ref{eq2.17}), we get the values of the exponent by numerical calculations in Figure \ref{fig:1}. In the figure, the bound is violated in certain ranges of the angular momentum which corresponds to specific spatial regions. The ranges of the angular momentum and spatial regions for the violation increase with the increase of the rotation parameter's value. For the fixed $a$ and $b$, the angular momentum's ranges where the bound is violated when the black hole and particle rotate in the opposite directions is larger than those when they rotate in the same directions. It is more likely to cause the violation when they rotate in the opposite directions. Due to different values of the rotation parameter, the values of the angular momentum corresponding to the maximum values of the exponent are different. There is $\lambda ^{2}-\kappa ^{2}>0$ when the angular momentum is zero, which means the black hole can violate the bound without depending on the angular momentum of the particle. The values of $\lambda^2-\kappa^2$ tend to constants when the angular momentum is large enough. When the rotation parameter is large enough and their rotation directions are opposite, the spatial region is relatively large. 

\begin{figure}[h]
	\centering
	\includegraphics[scale=1.4]{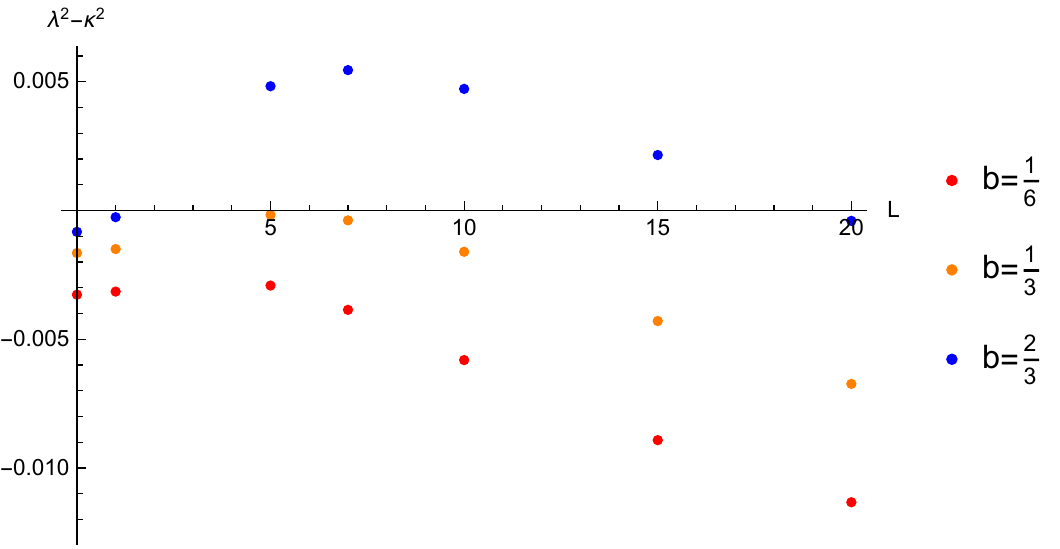}
	\caption{The influence of the angular momentum of the particle around the charged dilatonic black hole on the Lyapunov exponent. The bound on the Lyapunov exponent is violated in the range $1.22<L<19.16 (1.02042099r_{+}<r_{0}<1.45644077r_{+})$ when $b=\frac{2}{3}$. There is no violation when $b=\frac{1}{3}$ and $b=\frac{1}{6}$.}
	\label{fig:3}
\end{figure}

When $a= 0$, the metric (\ref{eq2.2}) describes a charged, non-rotating dilatonic black hole. To investigate the bound, we first derive the positions of equilibrium orbits for different values of $b$ and $L$, and then list them in Table \ref{t.2}. The horizon is located at $r_{+}=1.66667$ when $b=\frac{1}{6}$, at $r_{+}=1.33333$ when $b=\frac{1}{3}$, and at $r_{+}=0.66667$ when $b=\frac{2}{3}$. When the angular momentum increases, the location of $r_{0}$ is gradually far from the horizon. The influence of the angular momentum on the exponent is plotted in Figure \ref{fig:3}. In the figure, the violation occurs only for the certain values of the dilatonic constant and angular momentum. For example, the bound is violated in the range $1.22<L<19.16$ (the spatial region is $1.02042099r_{+}<r_{0}<1.45644077r_{+}$) when $b=\frac{2}{3}$. When the dilatonic constant is less than a certain value, there is no violation no matter how the angular momentum increases. When the value of the dilatonic constant is greater than a certain value, one can take a specific value of the angular momentum to violate the bound.

\begin{table}[h]
	\centering
	\begin{tabular}{cccccccc}
		\hline
		&L&0&1&5&7&10&15\\
		\hline
		\multirow{3}{*}{$r_{0}$}&$b=\frac{1}{6}$&1.69369&1.70104&1.82119&1.89166&1.98366&2.09891\\
		\cline{2-8}
		&$b=\frac{1}{3}$&1.34676&1.35150&1.43578&1.49030&1.56623&1.66821\\
		\cline{2-8}
		&$b=\frac{2}{3}$&0.67336&0.67807&0.75162&0.79400&0.85036&0.92421\\
		\hline
	\end{tabular}
	\caption{Locations of equilibrium orbits of the particle around the charged dilatonic black hole are gotten for different values of the dilatonic constant. }
	\label{t.2}
\end{table}

\begin{figure}[h]
	\centering
	\includegraphics[scale=1.5]{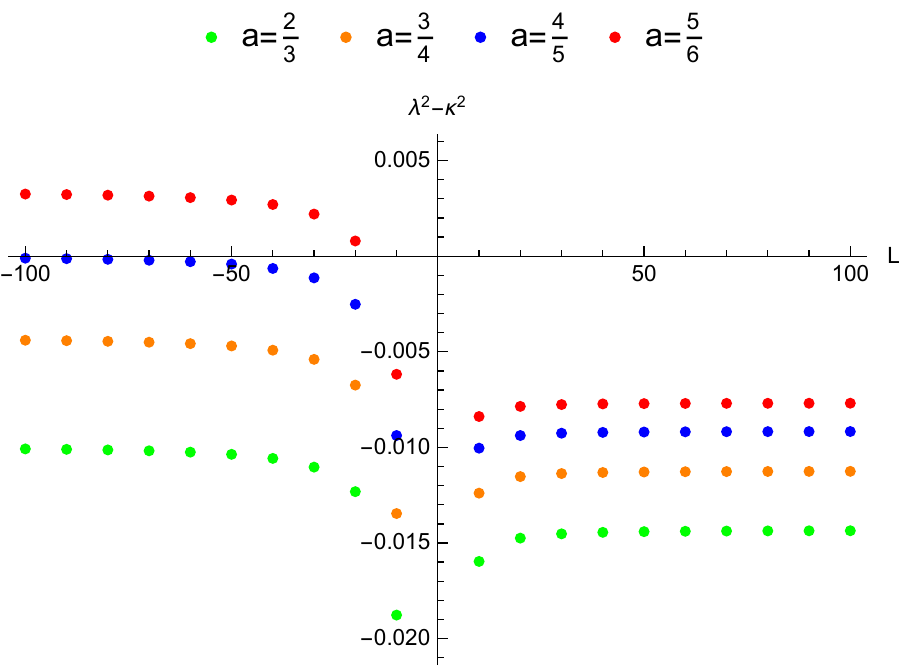}
	\caption{ The influence of the angular momentum of the particle around the non-extremal Kerr black hole on the Lyapunov exponent. The bound on the Lyapunov exponent is violated in the range $L<-17.40(2.47909864r_{+}<r_{0}<2.52593108r_{+})$ when $a=\frac{5}{6}$.}
	\label{fig:2}
\end{figure}

When $b = 0$, the metric (\ref{eq2.2}) is reduced to the Kerr metric. The values of the exponent of the chaos for a neutral particle around the Kerr black hole is plotted in Figure \ref{fig:2}. In the figure, when the rotation parameter and angular momentum are large enough, and the rotation directions of the particle and black hole are opposite, the bound is violated by increasing the angular momentum. The violation occurs in the range $L<-17.40(2.47909864r_{+}<r_{0}<2.52593108r_{+})$ when $a=\frac{5}{6}$. Although the range of the angular momentum where the bound is violated is large, the corresponding spatial region is not large. There is no violation for the bound when the rotation parameter is less than a certain value, or when the particle and black hole rotate in the same direction.  

\subsection{Lyapunov exponent in extremal EMDA black holes}

\begin{figure}[h]
	\centering
	\begin{minipage}[t]{0.48\textwidth}
		\centering
		\includegraphics[scale=0.64]{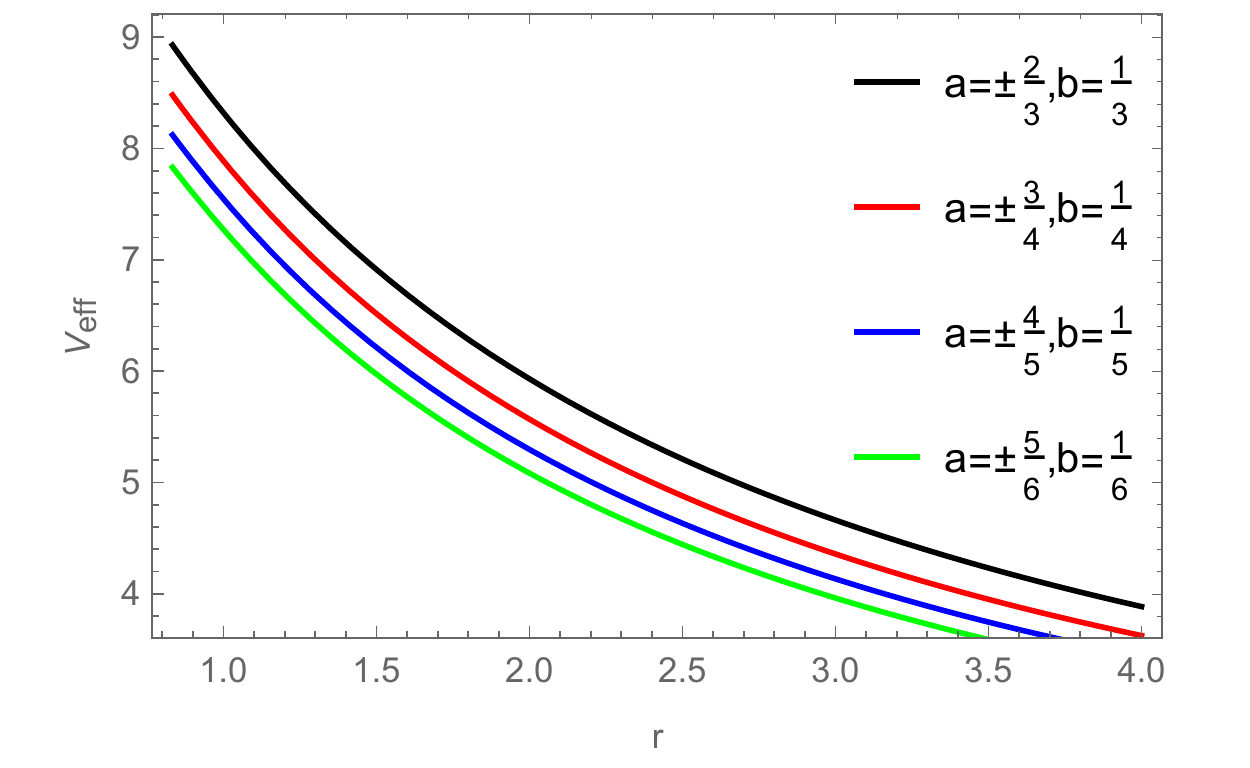}
	\end{minipage}
	\begin{minipage}[t]{0.48\textwidth}
		\centering
		\includegraphics[scale=0.58]{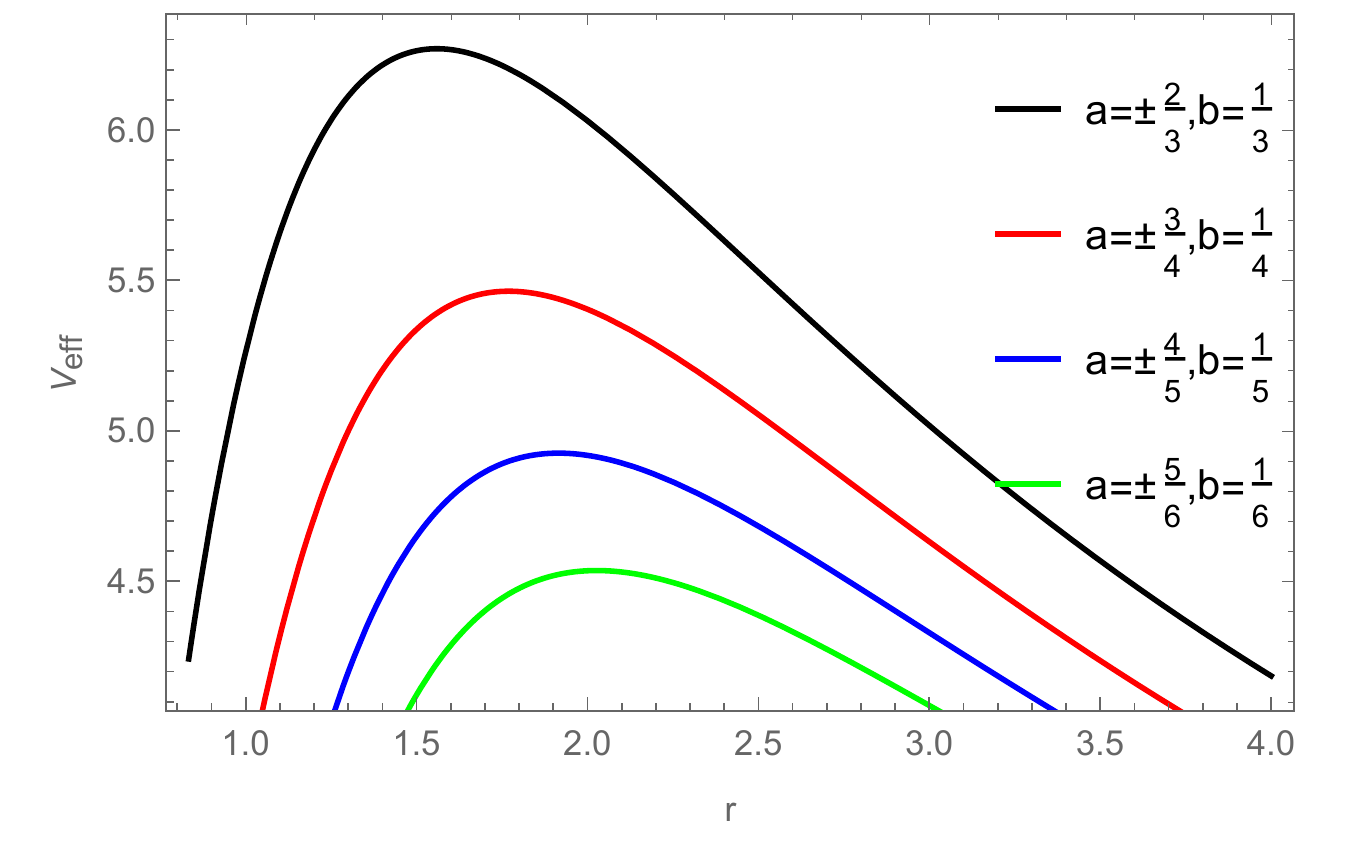}
	\end{minipage}
	\caption{The effective potentials at different positions outside the extreme EMDA black hole are plotted, where  $L=\pm 7$. The cases that the particle and black hole rotate in the same directions is plotted in the left figure, and that they rotate in opposite directions is plotted in the right figure.}
	\label{fig:4}
\end{figure}

For an extremal EMDA black hole, the inner and event horizons coincide with each other, and the surface gravity is zero. From Eq. (\ref{eq2.5}), we get $r_+= M_{0}$ and $a=\pm M_{0}$. Here we also let $M=1$, $m=1$ and $q=15$. Since the value of $\left.\sqrt{h}f\right|_{r=r_{0}}$ in Eq. (\ref{eq2.17}) is always positive, the sign of the value of the Lyapunov exponent depends on that of ${V}''_{eff}(r)$. We evaluate the violation of the bound by the positive and negative values of ${V}''_{eff}(r)$. The appearance of the maximum of the effective potential implies that ${V}''_{eff}(r)$ is less than zero. Using Eq. (\ref{eq2.14}), we get the effective potential at different positions in Figure \ref{fig:4}. In this figure, the potential has maximum values for different $a$ and $b$, which leads to ${V}''_{eff}(r)<0$. 

\begin{figure}[h]
	\centering
	\includegraphics[scale=1.5]{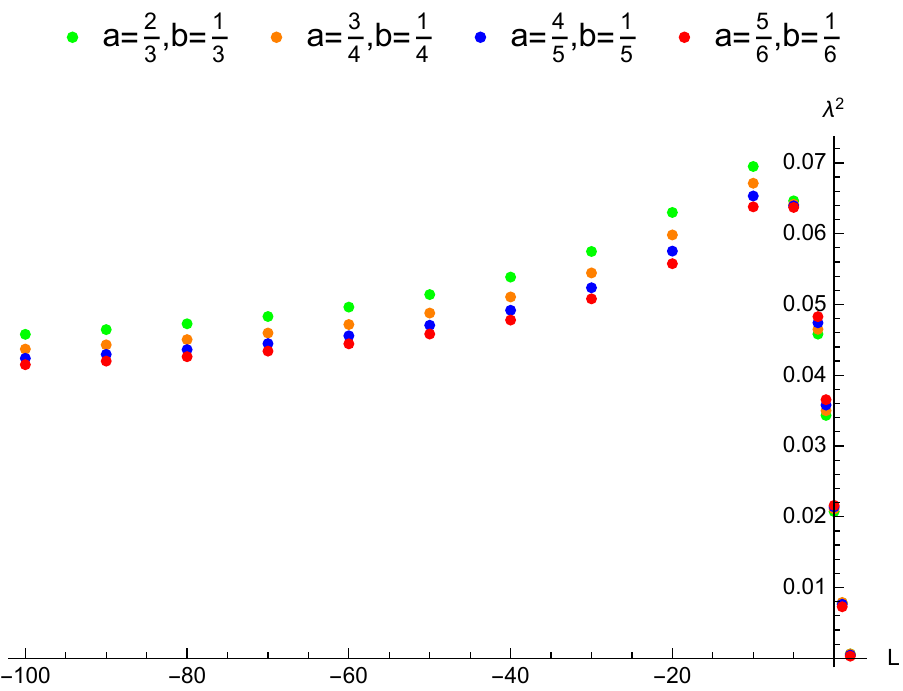}
	\caption{The influence of the angular momentum of the particle around the extremal EMDA black hole on the Lyapunov exponent.  The bound on the Lyapunov exponent is violated in the range $L<2.54(1.00304200r_{+}<r_{0}<4.44949278r_{+})$ when $a=\frac{2}{3}$, violated in the range $L<2.52(1.00756435r_{+}<r_{0}<4.30940108r_{+})$ when $a=\frac{3}{4}$, violated in the range $L<2.49(1.00119340r_{+}<r_{0}<4.23606798r_{+})$ when $a=\frac{4}{5}$, and violated in the range $L<2.38(1.00577200r_{+}<r_{0}<4.19089368r_{+})$ when $a=\frac{5}{6}$.}
	\label{fig:5}
\end{figure}

To investigate the influence of the angular momentum of the particle around this extremal black hole on the exponent, we draw Figure \ref{fig:5}. In the figure, the angular momentum's range and spatial region decrease with the increase of the rotation parameter when the bound is violated. The bound is always violated when the particle and black hole rotate in the opposite directions. The values of the exponent approach positive constants when the angular momentum is very large. The spatial regions where the bound is violated for the extremal EMDA black hole are significantly larger than those for the non-extremal EMDA black hole. One reason is the disappearance of the surface gravity for the extremal black hole.

\begin{figure}[h]
	\centering
	\begin{minipage}[t]{0.49\textwidth}
		\centering
		\includegraphics[scale=0.64]{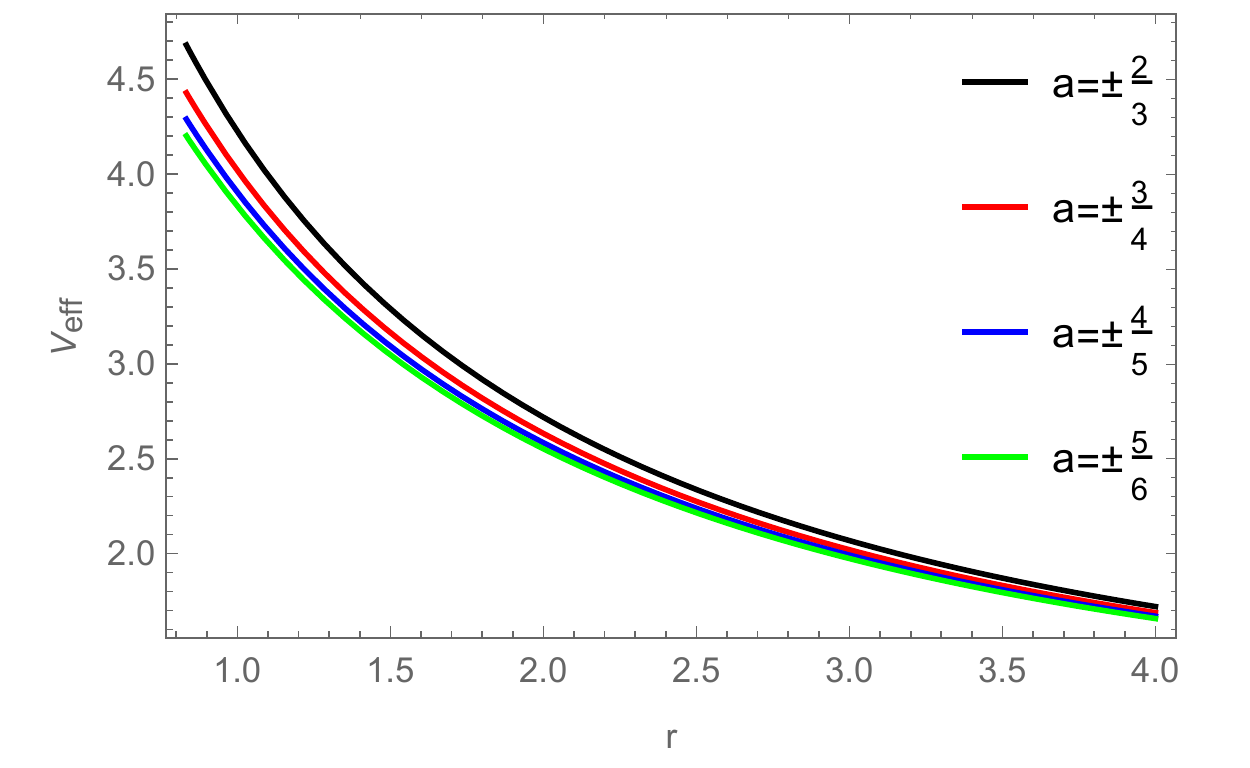}
	\end{minipage}
	\begin{minipage}[t]{0.49\textwidth}
		\centering
		\includegraphics[scale=0.64]{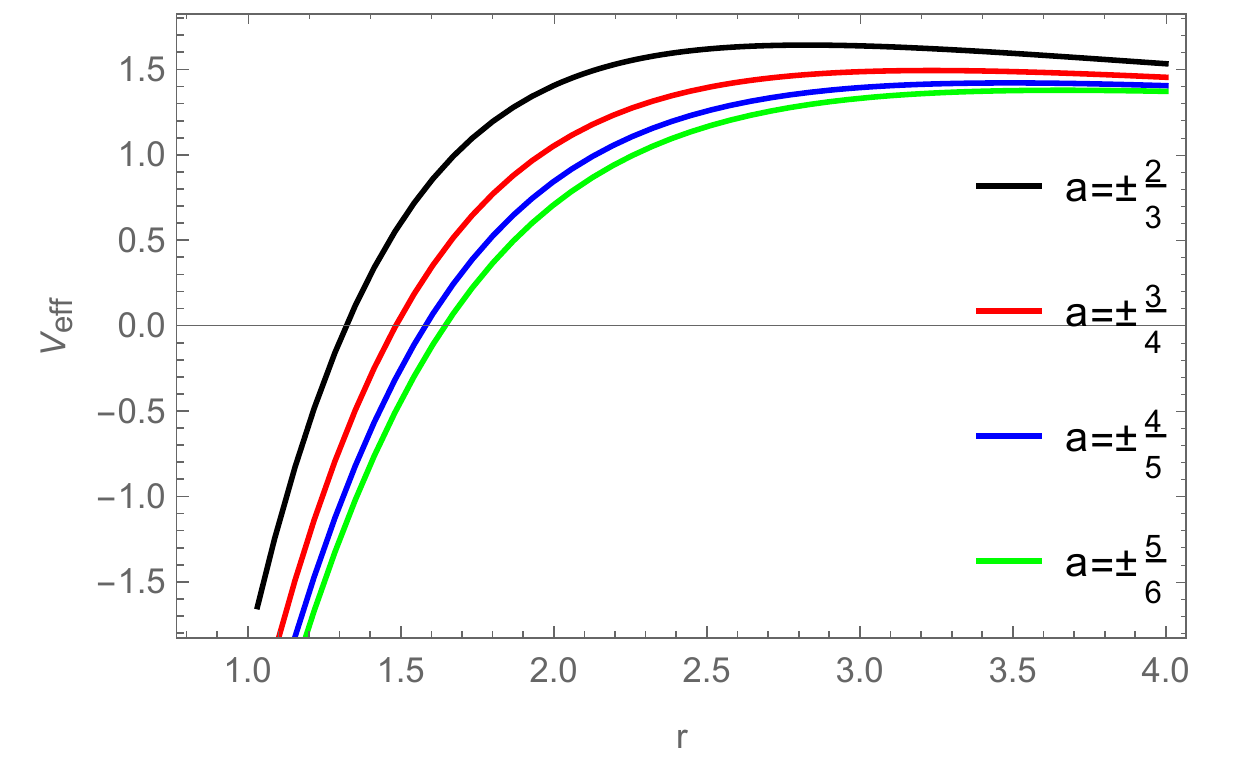}
	\end{minipage}
	\caption{The effective potential of the particle at different positions outside the extreme Kerr black hole, where $L=\pm 7$. $L$ and $a$ have the same signs in the left picture and different signs in the right picture.}
	\label{fig:6}
\end{figure}

Kan and Gwak have studied the Lyapunov exponent of the chaos of the particle around the extremal Kerr black hole in \cite{2022PhRvD.105b6006K}, where the rotation parameter took several specific values. Here we simply discuss the violation of the bound by taking into account several different values of the rotation parameter. Now $Q = 0$ in Eq. (\ref{eq2.14}) and the effective potential is plotted in Figure \ref{fig:6}. In this figure, there are maximum values in the effective potential for different values of the rotation parameter, which indicates the violation of the bound. This result is consistent with that gotten by them.

\subsection{Lyapunov exponent in near-horizon regions of EMDA black holes}

In this subsection, we investigate the exponent of the chaos of a particle in the near-horizon regions of the non-extremal and extremal EMDA black holes. We first focus our attention on the non-extremal black hole, and consider that the location of an equilibrium orbit is very close to the horizon. Let  

\begin{eqnarray}
r_{0}=r_{+}+\epsilon,
\label{eq3.1}
\end{eqnarray}

\noindent where $0<\epsilon \ll r_{+}$. Inserting the above relation into Eq. (\ref{eq2.15}) yields 

\begin{eqnarray}
{V}'_{eff}(r_{+}+\epsilon)
&=&\frac{\sqrt{r^2_{+}-r_{+}r_{-}}}{2}\frac{\sqrt{m^{2}(r_{+}+2b)(r_{+}^{2}+2br_{+}+a^{2})^{2}+r_{+}[aqQ-L(r_{+}+2b)]^{2}}}{(r_{+}^{2}+2br_{+}+a^{2})^{2}}\epsilon ^{-\frac{1}{2}}\nonumber\\
&+&\frac{1}{(r_{+}^{2}+2br_{+}+a^{2})^{2}}\left[\left(\frac{a^{2}(r_{+}-r_{-})}{r_{+}^{2}+2br_{+}+a^{2}}-2(r_{+}+b)\right)(aL+qQr_{+})\right.\nonumber\\
&+&\left.qQ\left(r_{+}^{2}+2br_{+}+a^{2}\right)-aL(r_{+}-r_{-})\right]+O(\epsilon ^{\frac{1}{2}}).
\label{eq3.2}
\end{eqnarray}

\noindent Without loss of generality, we let

\begin{eqnarray}
\frac{q}{L}=\frac{r_{+}+2b}{aQ}.
\label{eq3.3}
\end{eqnarray}

\noindent Using Eqs. (\ref{eq3.2}) and (\ref{eq3.3}) and ordering ${V}'_{eff}(r_{+}+\epsilon)=0$ yield  

\begin{eqnarray}
L=\frac{ma}{2}\sqrt{(r_{+}+2b)(r_{+}-r_{-}){r^{-1}_{+}}}\epsilon ^{-\frac{1}{2}}+\mathcal{O}(\epsilon ^{\frac{1}{2}}),
\label{eq3.4}
\end{eqnarray}

\noindent where $\mathcal{O}(\epsilon ^{\frac{1}{2}})$ contains the higher order terms of $\epsilon$ and is neglected. In the above relation, when the angular momentum is large enough, $\epsilon$ is very small and $r_0$ is very close to the horizon. It indicates that the assumption Eq. (\ref{eq3.1}) makes sense. Thus, the exponent in the near-horizon region is gotten as follow,

\begin{eqnarray}
\lambda^{2}-\kappa ^{2}=\frac{3(r_{+}-r_{-})}{8(r_{+}^{2}+2br_{+}+a^{2})^{2}}\epsilon +O(\epsilon ^{2}).
\label{eq3.5}
\end{eqnarray}

\noindent It is clearly that $\lambda^{2}>\kappa ^{2}$, which shows that the bound is violated in the near-horizon region of the non-extremal EMDA black hole. 

When $a= 0$, we use Eq. (\ref{eq2.15}) and get

\begin{eqnarray}
{V}'_{eff}(r_{+}+\epsilon )=\frac{\sqrt{r_{+}-r_{-}}\sqrt{m^{2}(r_{+}^{2}+2br_{+})+L^{2}}}{2(r_{+}^{2}+2br_{+})}\epsilon ^{-\frac{1}{2}}-\frac{qQ}{(r_{+}+2b)^{2}}+O(\epsilon ^{\frac{1}{2}})
\label{eq3.6}.
\end{eqnarray}

\noindent This implies that when the angular momentum and charge of the particle take certain values, one can get a small value of $\epsilon$. Now the exponent is

\begin{eqnarray}
\lambda^{2}-\kappa ^{2}=\frac{8(r_{+}+b)[m^{2}(r_{+}^{2}+2br_{+})+L^{2}]- r_{+}(r_{+}+3)(r_{+}+2b)}{2r_{+}(r_{+}+2b)^{3}[m^{2}(r_{+}^{2}+2br_{+})+L^{2}]}\epsilon +O(\epsilon ^{2}).
\label{eq3.7}
\end{eqnarray}

\noindent It is not difficult to prove that the first term at the right hand of the equal sign is always larger than zero in the large-$L$ limit, which leads to the violation of the bound in the near-horizon region of the charged dilatonic black hole. 

When the EMDA black hole is extremal, $\Delta =\epsilon ^{2}$ is obtained from Eqs. (\ref{eq2.4}) and (\ref{eq3.1}). We insert this relation into Eq. (\ref{eq2.15}) and get

\begin{eqnarray}
{V}'_{eff}(r_{+}+\epsilon )&=&\frac{\sqrt{Hr_{+}}-2aL(r_{+}+b)}{4r^2_{+}(r_{+}+b)^{2}}-\frac{1}{8r^3_{+}(r_{+}+b)^{3}}\left[\frac{2r_{+}^{2}(r_{+}+b)}{\sqrt{Hr_{+}}}(4m^{2}br_{+}(r_{+}+b)^{2}\right.\nonumber\\
&+&\left.(aqQ-L(r_{+}+2b))(aqQ-2bL))-2ar_{+}(aqQ-L(r_{+}+2b))\right]\epsilon \nonumber\\
&+&\mathcal{O}(\epsilon ^{2}).
\label{eq3.8}
\end{eqnarray}

\noindent It is obviously that there is a solution in the above equation when $\epsilon $ is very small. From Eq. (\ref{eq2.17}), we get

\begin{eqnarray}
\lambda ^{2}=\frac{2\sqrt{Hr_{+}}\mu +(r_{+}^{2}+2br_{+}+a^{2})(\rho +\jmath )}{r_{+}(r_{+}^{2}+2br_{+}+a^{2})^{3}H}\epsilon ^{3}+\mathcal{O}(\epsilon ^{4}),
\label{eq3.9}
\end{eqnarray}

\noindent where

\begin{eqnarray}
H&=&m^{2}(r_{+}+2b)(r_{+}^{2}+2br_{+}+a^{2})^{2}+r_{+}[aqQ-L(r_{+}+2b)]^{2},\nonumber\\
\mu&=&aL(6r_{+}^{2}+12br_{+}+4b^{2}+a^{2})+qQ[r_{+}^{2}(3r_{+}+4b)-2a^{2}(r_{+}+b)],\nonumber\\
\rho&=&m^{2}(r_{+}^{2}+2br_{+}+a^{2})[(r_{+}^{2}+2br_{+}+a^{2})(3r_{+}^{2}+8br_{+}+4b^{2}+a^{2})r_{+}\nonumber\\
&-&(2r_{+}+1)(r_{+}+2b)],\nonumber\\
\jmath &=&2r_{+}^{2}[aqQ-(r_{+}+2b)L][(r_{+}+2b-1)L-aqQ].
\label{eq3.10}
\end{eqnarray}

\begin{figure}[h]
	\centering
	\includegraphics[scale=1.1]{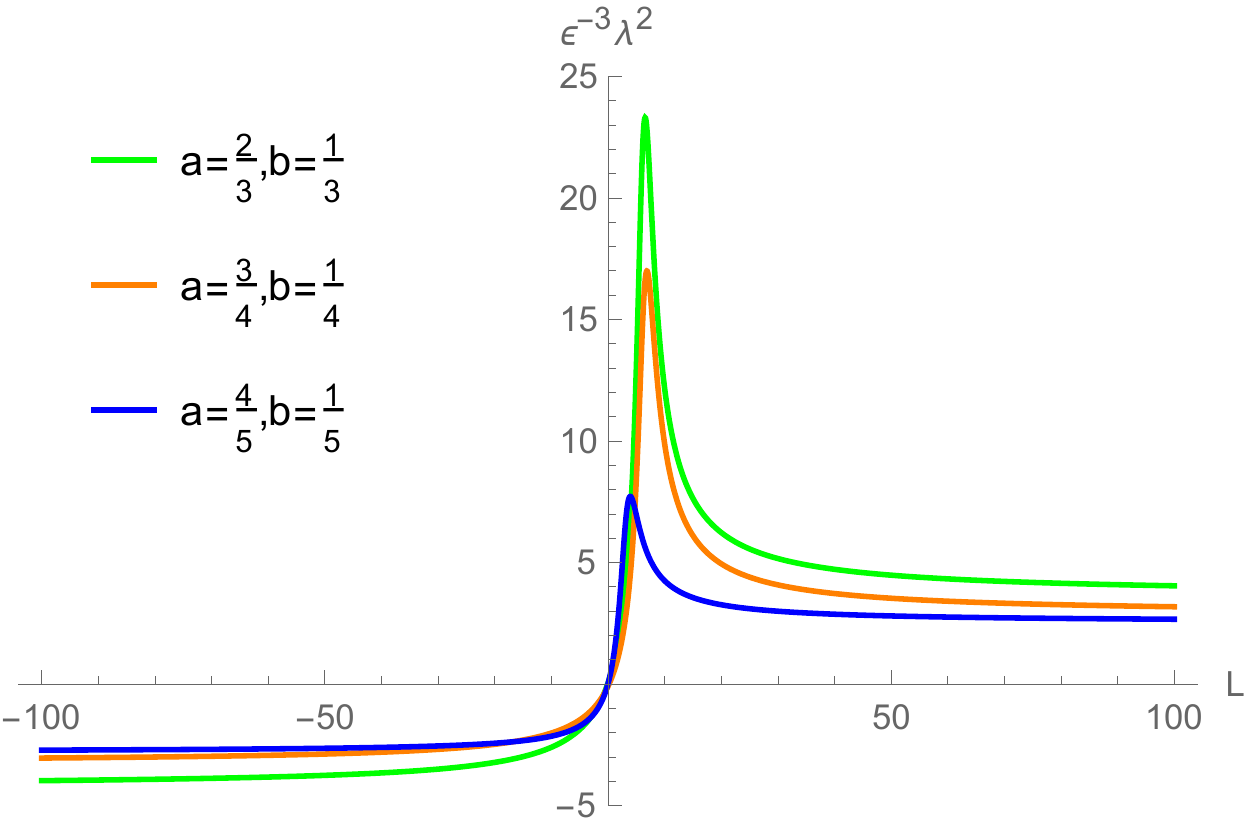}
	\caption{The influence of the angular momentum of the particle in the near-horizon region of the extremal EMDA black hole on the Lyapunov exponent.}
	\label{fig:7}
\end{figure}

We use numerical calculations to evaluate whether the coefficient of $\epsilon ^{3}$ in Eq. (\ref{eq3.9}) is larger than zero. In Figure \ref{fig:7}, there is $\lambda^2<0$ when $a>0$ and $L<0$, which shows that the bound is satisfied in the near-horizon region when the particle and black hole rotate in the opposite directions. When $a>0$ and $L>0$, the values of the exponent are always larger than zero. This implies that there is always a violation for the bound when the particle and black hole rotate in the same directions, and the angular momentum only affects the value of the exponent.

\section{Conclusion and discussion}\label{c}

In this paper, we investigated the influence of the angular momentum of the charged particle around the non-extremal and extremal EMDA black holes on the Lyapunov exponent. The angular momentum's range and spatial region where the bound of the exponent is violated were found. For the non-extremal black hole, the bound is violated when the dilatonic constant is fixed at $\frac{1}{3}$ and $a=\frac{2}{7}$, $\frac{1}{3}$, $\frac{2}{5}$ or $\frac{1}{2}$. The spatial regions for the violation increase with the increase of the value of the rotation parameter $a$. For the extremal black hole, the violation was also found when $a=\frac{2}{3}$, $\frac{3}{4}$, $\frac{4}{5}$ and $\frac{5}{6}$. The angular momentum's range and spatial region decrease with the increase of the rotation parameter when the bound is violated. The bound is always violated when the particle and black hole rotate in the opposite directions. It is more likely to cause the violation when the particle and black holes rotate in the opposite directions. The spatial regions where the bound is violated for the extremal black hole are relatively larger than those for the non-extremal black hole. In the near-horizon regions, there always exists the violation for the non-extremal black hole when the angular momentum is very large. The violation occurs when the particle and the extremal black hole rotate in the same directions.

The violation for the bound in the Kerr-Newman and Kerr-Newman AdS black holes was studied in \cite{2022PhRvD.105b6006K,2022arXiv220307298G}. For the non-extremal Kerr-Newman black holes, the authors found that the bound is violated when the particle and black holes rotate in the opposite directions. The bound is also violated in the near-horizon region when the angular momentum of the particle is very large. For the extremal Kerr-Newman black holes, there are violations when the particle and black holes rotate in the opposite directions, or the rotation parameter and black holes' charge take different signs. In \cite{2022arXiv220307298G}, they found that the negative cosmological constant reduces the chaotic behavior of the particle. In our work, the violation occurs within certain ranges of the angular momentum when the particle and non-extremal (or extremal) black holes rotate in the opposite directions. In the near-horizon regions, the violation occurs when the particle and extremal black hole rotate in the same directions, and doesn't occur when they rotate in the opposite direction. 

In this paper, although we got the violation for the bound, this violation may be not contrary to the conjecture in \cite{2016JHEP...08..106M}. Because they conjectured the upper bound of the exponent in the general thermal quantum systems with a large number of degrees of freedom. While we investigated the bound by using the motion of a single particle outside the horizon. As elaborated in \cite{2018JCAP...12..036L}, this result does not necessarily show that the bound conjectured in \cite{2016JHEP...08..106M} is violated.

\acknowledgments
This work is supported by the NSFC (Grant No. 12105031) and Tianfu talent project.

\end{document}